\documentclass{kluwer}    

\newdisplay{guess}{Conjecture}
\usepackage{epsfig}

\begin{document} 
\begin{article}
\begin{opening}         
\title{A self-consistent determination of the temperature profile and
  the magnetic field geometry in winds of late-type stars}   
\author{A. A. \surname{Vidotto}} 
\author{D. \surname{Falceta-Gon{\c c}alves}}
\author{V. \surname{Jatenco-Pereira}}
\runningauthor{Vidotto, Falceta-Gon{\c c}alves and Jatenco-Pereira}
\runningtitle{Alfv{\'e}n waves driven winds of late-type stars}
\institute{Depto. de Astronomia - Univ. de S{\~a}o Paulo - Brazil
\email{aline@astro.iag.usp.br}}
\begin{ao}
Depto. de Astronomia, Univ. de S{\~a}o Paulo, Rua do Mat{\~a}o, 1226,
S{\~a}o Paulo, SP, 05508-900, Brazil
\end{ao}
\date{\today}

\begin{abstract}
Cool giant and supergiant stars generally present low velocity winds
with high mass loss rates. Several models have been proposed to
explain the acceleration process of these winds. Although dust is
known to be present in these objects, the radiation pressure on these
particles is uneffective in reproducing the observed physical
parameters of the wind. The most promising acceleration mechanism
cited in the literature is the transference of momentum and energy
from Alfv{\'e}n waves to the gas. Usually, these models consider the wind
to be isothermal. We present a stellar wind model in which the Alfv{\'e}n
waves are used as the main acceleration mechanism, and determine the
temperature profile by solving the energy equation taking into account
both the radiative losses and the wave heating. We also determine
self-consistently the magnetic field geometry as the result of the
competition between the magnetic field and the thermal pressures
gradient. As main result, we show that the magnetic geometry present a
super-radial index in the region where the gas pressure is
increasing. However, this super-radial index is greater than that
observed for the solar corona.  
\end{abstract}
\end{opening}

\section{Introduction}
After decades of theoretical and observational studies of cool giant
and supergiant stars, the mechanisms by which the wind acceleration
occurs are still poorly understood. Compared to the Sun, these stars
are known to present 
continuous mass loss process occuring at high rates, typically
$10^{-10}-10^{-5}$~M$_{\odot }$~yr$^{-1}$, but in low velocity winds
($u_{\infty }<300$~km~s$^{-1}$) (Dupree 1986, Lamers \& Cassinelli
1999). Radiative pressure on grains 
transfers momentum to these particles being responsible 
for their acceleration and, if gas and dust are dynamically well coupled,
grains drag the gas outwards resulting in the mass ejection. However, for
stationary envelopes ($e.g.$ pre-AGB phase) dust driven theoretical models
have lately failed in reproducing the wind properties, mainly because the
dust-gas coupling is not effective (Sandin \& Hofner 2003). Observationally,
Guandalini $et$ $al.$ (2005) found no strong correlation between the mass
loss rates and the luminosities of AGB stars. Their main conclusion is that,
if radiative pressure is important in powering these stellar winds, it must
occur in addition to other mechanism. Another drawback to the radiation
pressure models is the need for the dust formation region to be close to the
star. Recent high resolution Doppler measurements show that winds are mainly
accelerated near the stellar surface ($r<1.3~ R_{\star}$) (Airapetian,
Carpenter \& Ofman 2003), while grains are expected to grow and survive at
even larger distances.

In this sense, another mechanism must be used to accelerate the gas near
surface. The most promising mechanism for the winds of cool stars is the
transference of momentum and energy to the wind from MHD waves. Hartmann \& MacGregor (1980) showed that it
would be possible to reproduce the observed low wind velocities and the high
mass loss rates of the cool giant and supergiant stars if some kind of wave
damping mechanism is effective at the wind basis ($r < 2$ R$_\star$). Jatenco-Pereira \& Opher (1989) studied the effects of
different damping mechanisms and magnetic field divergence and 
they showed that the magnetic field divergent geometry can rapidly dilute
the wave flux and also slow down the wind. Their magnetic field geometry was
based on empirical relations found from observations of the solar wind. This
because in general, in lack of direct measurements of the magnetic field
fluctuations and structure in other stars, we have to simply extrapolate our
knowledge from solar observations.

In this work, we model the acceleration and heating of a late-type
supergiant stellar wind considering an outwarded flux of Alfv\'en waves. We
solve the MHD equations to, self-consistently, determine the magnetic field
geometry and the wind temperature and velocity profiles. In section
2, we describe the model basic equations. In
section 3 we present the results and the discussions, followed, by 
the work conclusions.

\section{The Model}
The basic wind equations are based on mass, momentum, energy and magnetic
flux conservation. The first is given by $\rho u A(r) = \rho_0 u_0
A(r_0)$, where $u$ is the flow velocity, $\rho$ is the gas density and
$A(r)$ is the flow cross-section area at a distance $r$ from the
center of the star. The index ``$0$" indicates the variable is being
evaluated at the stellar surface.

Assuming a steady flow, the momentum equation can be written as: 
\begin{eqnarray}  \label{momentum}
\rho (\vec{u}\cdot \vec{\nabla })\vec{u} = - \frac{\rho GM_{\star
}}{r^{3}}\vec{r}-\vec{\nabla}\left(P + \frac{\langle (
  \delta B) ^{2}\rangle }{8\pi }\right) + \frac{1}{4\pi} (\vec{\nabla}
\times \vec{B}) \times \vec{B} ,
\end{eqnarray}
where $P = \rho k_B T / m$ is the thermal pressure, $k_B$ is the Boltzmann
constant, $T$ is the gas temperature, $m$ is the mean mass per particle, $G$
the gravitational constant and $\delta B$ the wave magnetic field amplitude.
In Equation (\ref{momentum}), the right hand side contains the
gravitational force 
density and the thermal and wave pressure gradients, respectively. The last
term represent the magnetic force. The wave amplitude ($\delta B$) is
related to the wave energy density ($\epsilon$) by $\epsilon = \langle
(\delta B)^2 \rangle/ (4 \pi)$.

\subsection{Thin fluxtube approximation}
Typically, considering magnetic field strengths $> 1$ G, the wind basis is
characterized by the relation $\beta = P/(B^{2}/8\pi) \ll 1$, $i.e.$ the
plasma is magnetically dominated. In this case, if we assume the wind to be 
initiated at funnels anchored at the stellar surface, which are surrounded
by a plasma with lower magnetic field strength, the magnetic pressure inside
will push the gas and the funnel field lines will expand. The funnel
cross-section radius ($\mathcal{R}$) will grow super-radially up to a limit
value ($\mathcal{R}_m$). This limiting cross-section radius depends both on
the relation between external and internal magnetic field strengths and on
the filling factor ($\alpha$). As the area increases, the internal magnetic
strength diminishes until the equilibrium between internal and external
magnetic pressures is reached. If the internal magnetic field strength is
much larger than the external, the flux tubes cross-section will depend on
the filling factor only. The filling factor is the ratio between the area of
the stellar surface covered by funnels and the total area. The averaged
maximum area that a funnel could reach would be $A_m = A(r_0)/\alpha$ or, in
terms of the cross-section radius: $ \mathcal{R}_m = {\mathcal{R}_0} /
{\alpha^{1/2}}$. For the quiet Sun, the funnels that merge to form the
coronal holes cover about 10\% of the total surface.

In this work, at the wind basis, we assume the plasma to be magnetically
dominated and the left hand side of Equation (\ref{momentum}) may be
neglected if 
compared to the other terms. Then, by using $\vec{\nabla} \cdot 
\vec{B} = 0$ and a power series expansion method proposed by
Pneuman, Solanki \& Stenflo (1986), we can determine self-consistently the
magnetic field geometry without assuming any empirical function for the
funnel cross-section with distance.

Following Pneuman, Solanki \& Stenflo (1986), using the thin fluxtube
approximation, Equation (\ref{momentum}) and $\vec{\nabla} \cdot 
\vec{B} = 0$ are described near stellar surface, which in our model
occurs up to $\sim 1.4~r_0$, by: 
\begin{equation}\label{a}
4\pi \frac{\partial P}{\partial y} \simeq B_{r}\left( \frac{\partial B_{y}}{
\partial r}-\frac{\partial B_{r}}{\partial y}\right),
\end{equation}
\begin{equation}\label{b}
4\pi \left( \frac{\partial P}{\partial r}+\frac{P}{H}\right) \simeq -B_{y
}\left( \frac{\partial B_{y }}{\partial r}-\frac{\partial B_{r}}{\partial y }
\right),
\end{equation}
and 
\begin{equation}\label{c}
\frac{1}{y }\frac{\partial }{\partial y }\left( y B_{y }\right) +\frac{
\partial B_{r}}{\partial r}=0,
\end{equation}
where $H=k_BTr^{2}/GmM_{\star}$ is the scale height. The thin fluxtube
approximation is reliable when the cross-section radius 
$\mathcal{R}$ is negligible compared to both the scale height of 
the external medium and any variation scales along the
tube (Spruit 1981; Longcope and Klapper 1997). 

Expanding all variables as power series in $y$ ($i.e.$ along the tube
radius), and neglecting terms of orders higher than 2, Equations (\ref{a}) $-$
(\ref{c}) give rise to a differential equation for the fluxtube cross-section: 
\begin{eqnarray} \label{d}
\frac{A(r_{0})}{2H_{0}^{2}}\left[ \frac{\partial ^{2}}{\partial r^{2}}\left( 
\frac{A(r_{0})}{A(r)}\right) -\frac{1}{2A(r)}\frac{\partial }{\partial r}
\left( \frac{A(r_{0})}{A(r)} \right)\right] =  \nonumber \\
\left( \frac{A(r_{0})}{A(r)} \right) ^{2} \left[ 1-\left( \frac{B_{\rm
      ext}}{
B_0}\frac{\left( 1-\alpha \right) }{\left( \frac{A(r_{0})}{A(r)}-\alpha
\right) }\right) ^{2}\right] +2\beta \frac{P(r)}{P(r_{0})},
\end{eqnarray}
where $\beta = 4 \pi P(r_0)/B_0^2$, $\alpha$ is the filling factor and
$B_{\rm ext}$ is the magnetic field strength outside the fluxtube. In
the following calculations we fixed its value to be $10^{-3}B(r)$.

To simplify the set of equations, we will define the funnel area expansion
as a function of radial distance by: $A\left( r\right) =A\left(
r_{0}\right) \left( {r}/{r_{0}}\right) ^{S}$, 
where $S$ is the expansion index, which is super-radial ($S > 2$) at the
wind basis up to the merging radius when $S$ becomes 2. $S$ is determined
from Equation (\ref{d}).

\subsection{The wind equations}
In a consistent model, to avoid assuming any empirical function for the 
magnetic field geometry and to
determine the wind temperature at each wind position ($r$), we have to solve
the energy equation, which is determined from the balance between wave
heating and the adiabatic expansion and radiative coolings (Hartmann,
Edwards \& Avrett 1982). Thus, neglecting conduction, we write the energy
equation as: 
\begin{equation}  \label{energy}
\rho u \frac{d}{dr} \left( \frac{u^2}{2} + \frac52 \frac{k_B T}{m} - \frac{G
M_\star}{r} \right) + \frac{u}{2} \frac{d \epsilon}{dr} = (Q - P_R) \, ,
\end{equation}
where $Q$ is the wave heating rate, $i.e.$ the rate at which the gas is
being heated due to dissipation of wave energy, and $P_R$ is the radiative
cooling rate, both in erg~cm$^{-3}$~s$^{-1}$. The wave heating can be
written as $Q = {\epsilon}(u + v_A)/ L$ and the radiative cooling is
given by $P_R = \Lambda \, n_e \, n_H$, where $n_e$ is the electron
density, $n_H$ is the hydrogen density and $\Lambda$ is the radiative
loss function. Here, we adopt the $\Lambda$ 
function given by Schmutzler \& Tscharnuter (1993) and calculate $n_e$ with
the modified Saha equation given by Hartmann \& MacGregor (1980).

The wave energy density at each step may be calculated using a WKB
approximation, from the wave action conservation. This approximation can
often be employed when the properties of the medium vary slowly on a
scale comparable to the wavelength (Usmanov {\it et al.} 2000). Under this
assumption, the wave energy density is dissipated as follows: 
\begin{equation}
\epsilon = \epsilon_0 \frac{M_0}{M} \left( \frac{1+M_0}{1+M} \right)^2 \exp 
\left[ - \int_{r_0}^{r} \frac{1}{L} \, dr^{\prime}\right] \, ,
\end{equation}
where $M=u/v_A$ is the Alfv{\'e}n-Mach number, $v_A = (B / \sqrt{4 \pi \rho})
$ the Alfv{\'e}n speed and $L$ the wave damping length. Also, the wave
flux ($\phi_A$) at $r_0$ is evaluated by $\phi_{A_{0}} = \epsilon_0
v_{A_{0}} \left( 1 + 1.5 M_{0} \right)$. We consider the non-linear
damping mechanism, which length is given by (Jatenco-Pereira \& Opher
1989):  
\begin{equation}
L = L_{0}\left( \frac{v_{A}}{v_{A_{0}}}\right) ^{4}\frac{\langle
(\delta v)^{2}\rangle _{0}}{\langle (\delta v)^{2}\rangle }
\left( 1+M\right) \, ,
\end{equation}
where $\langle(\delta v)^{2}\rangle$ is the averaged squared
perturbation velocity amplitude and $L_0$ is the
damping length at the wind basis, which is mainly dependent on the assumed
wave frequency spectrum (Lagage \& Cesarsky 1983). Here, we will let it as a
free parameter.

The equations presented here fully describe the wind parameters and
the magnetic field geometry under the given assumptions. In the next
section, we show the main results by applying these equations in a
typical cool supergiant star and compare them with previous works.

\section{Results and Discussions}
For the last decades, the validity of a wind model was limited to
simply reproduce the terminal velocity and the mass-loss rate of a
given star. As a consequence, a number of accelerating mechanisms were found
in accordance to observations. However, with the appearance of high
resolution observations from more sensitive intruments, in the near
future, the radial profiles of the wind parameters will become
measurable and will be decisive on the choice of the wind model. 

Here, we obtained the wind paramenter profiles applying the described
model on a cool supergiant star with $M_\star$ = 16 M$_\odot$, $r_0 =
400$ R$_\odot$, $\rho_0 = 10^{-13}$ g cm$^{-3}$, $B_0 = 10$ G and $T_0
= 3500$ K, as used by Hartmann \& MacGregor (1980) in their model star
4. We also assumed a filling factor $\alpha = 0.1$,
according to solar observations. For these stars, observational data
reveal typical mass-loss rates of $\dot{M} \simeq 10^{-7} - 10^{-6}$ 
M$_\odot$ yr$^{-1}$ and terminal velocities of $u \simeq 70$ km s$^{-1}$.
Unfortunately, the available data are limited in spatial resolution for most
of the stars, and it is not possible to fit the complete radial profiles.

\begin{figure}
\centerline{\includegraphics[scale=.35]{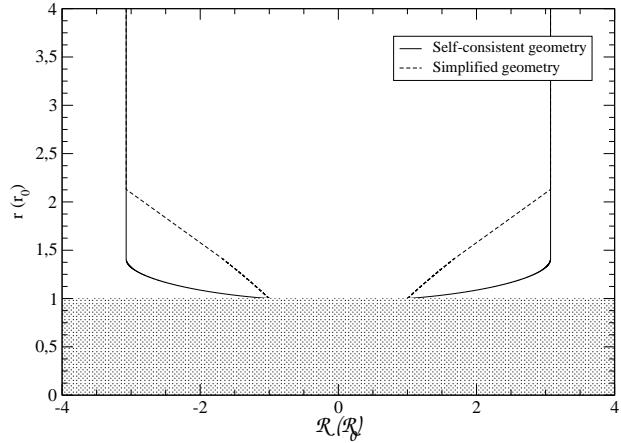}}
\caption{The magnetic field structure for a constant super-radial index 
$S = 5$ (dashed line) and that determined self-consistently in our
  model (solid line).}
\label{fig1}
\end{figure}

Assuming a surface magnetic field strength $B_0 = 10$ G, an Alfv\'en
waves flux of $\phi_{A0} = 10^7$ erg cm$^{-2}$ s$^{-1}$ at the wind
basis and a low damped wave flux ($L_0 = 5$ r$_0$), both the wind
terminal velocity and the mass loss rate obtained were
consistent with the observations. This value corresponds to a wave
amplitude of $\sqrt{\langle (\delta B)^2 \rangle} \simeq 3\times
10^{-2}~B_0$, which is very plausible for a turbulent medium as that
at the stellar surface (Suzuki \& Inutsuka 2005). 

\begin{figure}
\centerline{\includegraphics[scale=.35]{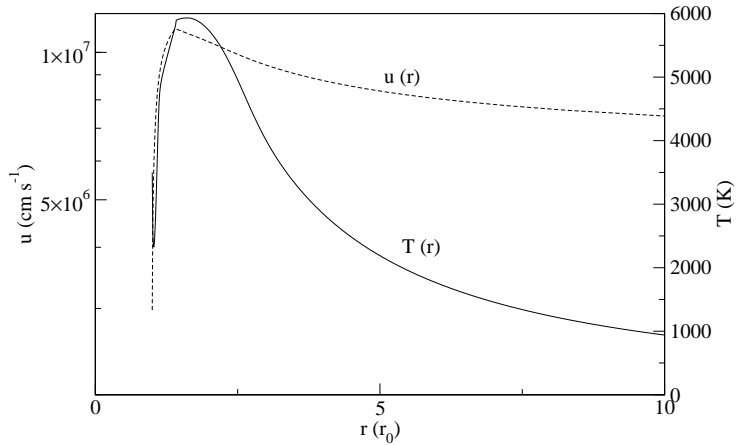}}
\caption{The wind velocity profile and the wind temperature
  profile for the best fitting parameters in the case of low
  damped waves.} 
\label{fig}
\end{figure}

The velocity and temperature profiles for this case are shown in Figure \ref{fig}. The velocity profile reveals a peak of $u > 100$ km s$^{-1}$ at $r < 2.0$ r$_0$, and slightly decreases for larger distances until
reaching the observed value. The temperature profile presents an initial
negative gradient reaching temperatures $T < 2500$ K in a narrow region.
Also, near $r = 1.5$ r$_0$ the temperature reaches the maximum value of $\sim 6000$ K. For $r > 3.0$ r$_0$, where the wave heating and the radiative
losses are low, the temperature decreases due to the adiabatic expansion.

\section{Conclusions}
We propose a self-consistent wind model to determine the 
parameters profiles for a supergiant late-type star. 
To determine the magnetic field geometry we used 
an expansion method over the wind physical parameters as proposed by Pneuman, 
Solanki \& Stenflo (1986). Near surface, the magnetic 
pressure inside the flux tubes are higher 
than that of the surrounding medium, forcing the field lines to curve. 
We found an initial super-radial expansion factor $S > 30$ at the wind
basis, much  higher than the value used by previous authors
(e.g. Jatenco-Pereira \& Opher 1989, Bravo \& Stewart 1997, Dobrzycka
{\it et al.} 1999) that included empirical relations to account for
the magnetic field geometry based on solar observations. 

Considering a supergiant late-type star, we obtained the wind velocity
and temperature profiles. Typically, Alfv\'en waves driven winds
result in high velocity winds ($u > 100$ km s$^{-1}$), unless some
strong wave damping mechanism takes place at the wind basis. We showed
that this conclusion is correct in the case of low divergent magnetic
field structures. In this work, the strong divergence is responsible
for a rapid wave spatial dilution near surface, resulting in a lower
wind velocity even for low damped waves.  

The obtained results were consistent to the typical mass-loss
rate ($\dot{M} \simeq 10^{-7} - 10^{-6}$ M$_\odot$ yr$^{-1}$) and the
terminal velocity ($u \simeq 70$ km s$^{-1}$) observed for these
objects. The velocity profile reveals 
an efficient acceleration at $r < 1.5~r_0$, reaching the maximum
value $\sim 100$ km s$^{-1}$. In this region the wind is mainly
accelerated by the wave energy density and the thermal pressure
gradients. Afterwards, the absence of the wave acceleration and the
cooling gas result in a decrease of the velocity to the observed
values. For the temperature, assuming a weakly damped wave flux, the
radiative losses and the expansion cooling are dominant near surface,
and the temperature gradient is initially negative. The temperature
falls to $\simeq 2500$ K in a sharp region and then, as density
decreases as the wind accelerates and the flux tube expands, it
increases up to $\simeq 6000$ K at $r < 2.0$ r$_0$. For higher
distances, where the radiative losses are low and the wave heating is
no longer effective, the temperature decreases mainly due to the
adiabatic expansion.  

Although the present calculations provided a new and interesting
picture of the physical processes involving the heating and the
acceleration mechanisms of cool stellar winds, the model presents a
limitation. The wind equations were solved considering the thin
fluxtube approximation. Mainly for a high divergence wind, a complete 
2-D set of equations must be employed in order to obtain more precise
results. This step should be concluded in the near future.

\section*{Acknowledgments}
A. A. Vidotto and D. Falceta-Gon\c calves thank FAPESP for the
finantial support (04/13846-6 and 04/12053-2) and V. 
Jatenco-Pereira thanks CNPq (304523/90-9).

\end{article}

\end{document}